\documentclass[twocolumn,showpacs,aps,prl,superscriptaddress,floatfix]{revtex4}
\usepackage{graphicx}
\usepackage{epsfig}
\usepackage{amsmath}
\usepackage{amssymb}
\usepackage{hhline}
\usepackage{multirow}
\usepackage{rotating}
\usepackage{fancyheadings}
\usepackage{relsize}
\RequirePackage{xspace}

\def\babar{\mbox{\slshape B\kern-0.1em{\smaller A}\kern-0.1em
    B\kern-0.1em{\smaller A\kern-0.2em R}}}

\def\cbar  {\ensuremath{\overline c}\xspace}
\def\B       {\ensuremath{B}\xspace}
\def\Bbar    {\kern 0.18em\overline{\kern -0.18em B}{}\xspace}

\def\BB      {\ensuremath{B\Bbar}\xspace} 
\def\Bz      {\ensuremath{B^0}\xspace}
\def\Bzb     {\ensuremath{\Bbar^0}\xspace}
\def\BzBzb   {\ensuremath{\Bz {\kern -0.16em \Bzb}}\xspace}
\def\Bu      {\ensuremath{B^+}\xspace}
\def\Bub     {\ensuremath{B^-}\xspace}

\def\BpBm    {\ensuremath{\Bu {\kern -0.16em \Bub}}\xspace}
\def\jpsi     {\ensuremath{{J\mskip -3mu/\mskip -2mu\psi\mskip 2mu}}\xspace}
\def\psitwos  {\ensuremath{\psi{(2S)}}\xspace}
\def\Y#1S{\ensuremath{\Upsilon{(#1S)}}\xspace}
\def\FourS {\Y4S}

\def\BR         {{\ensuremath{\cal B}\xspace}}
\newcommand{\mbtox}{\ensuremath{\B \to X e \nu}}

\newcommand{\mbtoxu}{\ensuremath{\B \to X_u e \nu}\ }
\newcommand{\mbtoxc}{\ensuremath{\B \to X_c e \nu}\ }

\newcommand{\gev}{\ensuremath{\mathrm{\,Ge\kern -0.1em V}}\xspace}
\newcommand{\mev}{\ensuremath{\mathrm{\,Me\kern -0.1em V}}\xspace}
\newcommand{\ev}{\ensuremath{\mathrm{\,e\kern -0.1em V}}\xspace}
\newcommand{\gevc}{\ensuremath{{\mathrm{\,Ge\kern -0.1em V\!/}c}}\xspace}
\newcommand{\mevc}{\ensuremath{{\mathrm{\,Me\kern -0.1em V\!/}c}}\xspace}
\newcommand{\gevcc}{\ensuremath{{\mathrm{\,Ge\kern -0.1em V\!/}c^2}}\xspace}
\newcommand{\mevcc}{\ensuremath{{\mathrm{\,Me\kern -0.1em V\!/}c^2}}\xspace}

\def\invfb   {\ensuremath{\mbox{\,fb}^{-1}}\xspace}

\def\ra                 {\ensuremath{\rightarrow}\xspace}
\def\to                 {\ensuremath{\rightarrow}\xspace}

\def\pep2{PEP-II}

\newcommand{\nimBaseA}       {Nucl.\ Instr.\ Meth.\xspace}
\newcommand{\nim}       [1]  {\nimBaseA~{\bf #1}}
\def\mynim  #1 #2 #3 {\nim{#1},\ #2 (#3)}
\newcommand{\progtp}    [1]  {{Prog.\ Theor.\ Phys.\ {\bf #1}}}

\def\pstar{\ensuremath{p^*}\xspace}
\newcommand{\emin}{\ensuremath{E_{0}}\xspace}
\newcommand{\ls}{\ensuremath{e^\pm e^\pm\ }}
\newcommand{\uls}{\ensuremath{e^+ e^-\ }}
\def\Ee    {\ensuremath{E_e}\xspace}
\def\epem       {\ensuremath{e^+e^-}\xspace}
\newcommand{\numbf}{10.36}
\newcommand{\numbfstaterr}{0.06}
\newcommand{\numbfsyserr}{0.23}

\begin{document}

\begin{flushleft}
\babar-PUB-03/045 \\
SLAC-PUB-10298\\
hep-ex/0403030 \\
\end{flushleft}

\title[Short Title]{\large \bf
Measurement of the Electron Energy Spectrum and its Moments\\ in Inclusive 
$B\to X e\nu$ Decays 
 }

%
\author{B.~Aubert}
\author{R.~Barate}
\author{D.~Boutigny}
\author{F.~Couderc}
\author{J.-M.~Gaillard}
\author{A.~Hicheur}
\author{Y.~Karyotakis}
\author{J.~P.~Lees}
\author{V.~Tisserand}
\author{A.~Zghiche}
\affiliation{Laboratoire de Physique des Particules, F-74941 Annecy-le-Vieux, France }
% ...
\collaboration{The \babar\ Collaboration}
\noaffiliation

\begin{abstract}
We report a measurement of the inclusive electron energy spectrum for 
semileptonic decays of \B mesons in a data sample of 52 million \Y4S 
$\to$ \BB decays collected with the \babar\ detector at the PEP-II 
asymmetric-energy $B$-meson factory at SLAC. We determine the branching 
fraction, first, second, and third moments of the spectrum for lower cut-offs 
on the electron energy between 0.6 and 1.5\gev. We measure the partial 
branching fraction to be
${\cal B}(B\to Xe\nu, E_{e}>0.6 \gev)= (\numbf \pm \numbfstaterr (\rm stat.) 
\pm \numbfsyserr (sys.)) \% $.
\end{abstract}

\pacs{12.15.Hh, 11.30.Er, 13.25.Hw} 

\maketitle

The operator product expansion provides corrections to the relation between
the semileptonic $B$ decay rate and the magnitude of the Cabibbo-Kobayashi-Maskawa
(CKM)~\cite{CKM} matrix element $V_{cb}$ in the free-quark model~\cite{theo}. The 
corrections are expressed in terms of non-perturbative quantities that can be 
extracted from moments of inclusive distributions. We plan to use the precision 
measurements of moments of the lepton energy spectra presented here and of hadron mass 
distributions~\cite{hadmom} to determine those parameters and thereby to improve the 
determination of $|V_{cb}|$~\cite{comb}.

In this paper, we present a new measurement of the inclusive electron energy spectrum 
from semileptonic $B$ decays, averaged over charged and neutral $B$ mesons produced 
at the \FourS\ resonance. After correcting for charmless semileptonic decays, we 
derive from this spectrum several moments as a function of the minimum electron 
energy ranging from 0.6 \gev to 1.5 \gev, where lower endpoint is set by the limits of 
electron identification and prevalence of background.
In the B meson rest frame, we define
$R_i(\emin,\mu)$ as $\int_{\emin}^{\infty} (\Ee-\mu)^i (d\Gamma/dE_e)\,dE_e$,
and measure the first moment $M_1(\emin) = R_1(\emin,0) / R_0(\emin,0)$,
the central moments $M_n(\emin)=R_n(\emin,M_1(\emin))/R_0(\emin,0)$ for $n$ = 2, 3 and the
partial branching fraction ${\cal B}(\emin) = \tau_B \; R_0(\emin,0)$, where 
$\tau_B$ is the average lifetime of charged and neutral $B$ mesons.

The measurements presented here are based on data collected  by the 
\babar\ detector~\cite{babar} at the PEP-II asymmetric $e^+e^-$ storage 
ring; they correspond to an integrated luminosity of 
47.4 \invfb\ on the \FourS\ resonance and 9.1 \invfb\ at an energy 
40 \mev below the resonance (off-resonance), measured in the electron-positron
center of mass frame. Where background and efficiency
corrections cannot be measured directly from data, we use a full simulation of 
the detector based on GEANT4~\cite{geant}. In the following, all kinematic
variables defined in the \FourS\ rest frame will be annotated with an
asterisk.

This analysis is similar to the \babar \  measurement of the semileptonic branching 
fraction~\cite{oldprl}, including use of the same electron identification criteria, but
supersedes it by an order of magnitude in integrated luminosity. We identify 
\BB\ events by observing an electron, $e_{tag}$, with charge $Q(e_{tag})$ and a momentum of 
$1.4<p^*<2.3~\gevc$ in the $\Upsilon(4S)$ rest frame. These electrons make up the 
tagged sample that is used as normalization for the branching fraction. 
Each electron $e_{sig}$ with charge $Q(e_{sig})$ for which we require $p^*>0.5 \gevc$ is assigned
to the unlike-sign sample if the tagged sample contains an electron
with $Q(e_{tag})=-Q(e_{sig})$, and to the like-sign sample if $Q(e_{tag})=Q(e_{sig})$.
In events without \BzBzb mixing, primary electrons from 
semileptonic \B decays belong to the unlike-sign sample while secondary electrons 
contribute to the like-sign sample.

Multi-hadron events are selected by either requiring a track multiplicity
$N_{ch}\ge 5$, or $N_{ch}=4$ plus at least two photon candidates with 
$E_{\gamma}>$ 80\mev. Track pairs from converted photons
are not included in $N_{ch}$, but count as one photon. For further 
suppression of non-\BB events we require the ratio of the Fox-Wolfram
moments $H_2^*/H_0^*$ to be less than 0.8.

Electrons originating from the same $B$ meson as the tagged electron
typically have opposite charge and direction. To reject them we require  
\begin{equation}
\cos\alpha^*   > 1.0 - p_e^*(\rm \gevc)  \ \ \rm{and} \ \   
\cos\alpha^*> -0.2, 
\label{oacut}
\end{equation}
where  $\alpha^*$ is the angle between the two electrons. This requirement
also excludes electron pairs from  $\jpsi\to e^+e^-$ decays.
To suppress background contributions from $\jpsi\to e^+e^-$ 
decays to the tagged sample, we require 
the invariant mass $M_{ee}$ of the tag electron, paired with any electron 
of opposite charge and  $\cos\alpha^*<-0.2$, to be outside the interval  
$2.9 < M_{ee} < 3.15 \gevcc$. Here the requirement on  $\cos\alpha^*$ does
not reduce the efficiency of this veto, but ensures that no signal electron
satisfying Eq.~\ref{oacut} is excluded from the unlike-sign sample.
The efficiencies of these selection criteria are estimated by Monte Carlo (MC) simulation.

Continuum background is subtracted from the tagged, like- and unlike-sign samples by scaling the 
off-resonance yields by the ratio of on- to off-resonance integrated 
luminosities, corrected for the energy dependence of the continuum 
cross section. In the off-resonance sample, the momenta are scaled by
the ratio of the on- and off-resonance energies.

Electron spectra from photon conversions and Dalitz decays are extracted 
from data, taking into account the pair-reconstruction efficiencies 
from MC simulation. The relative uncertainty in these efficiencies is 
estimated to be 13\% and 19\% for conversion and Dalitz pairs, respectively. 

The misidentification rates for pions, kaons, and protons are extracted 
from data control samples. They rise from $0.05\%$ to $0.12\%$ for pions 
and fall from $0.4\%$ to $0.1\%$ for kaons as $p^*$ increases from 0.5 to
2.5 \gevc. The systematic errors 
are estimated from the control sample purities and from the uncertainties 
in the $\pi$, $K$ and $p$ abundances. The resulting relative uncertainties 
are less than 40\%.

There is a small residual background in the sample of unlike-sign pairs originating 
from the same $B$ meson and fulfilling the requirement on the opening angle $\alpha^*$
from Eq.~\ref{oacut}. It is estimated from a fit to the $\cos\alpha^*$ 
distribution, separately for each 50-\mevc-wide bin in $p^*$. The distribution is flat for 
signal pairs, while for background pairs it is taken from MC simulation, with a 
maximum at $\cos\alpha^*=-1$ and gradually decreasing to 0 at $\cos\alpha^*=1$. 

\begin{figure}[h]
\begin{center}
\includegraphics[height=3.2in]{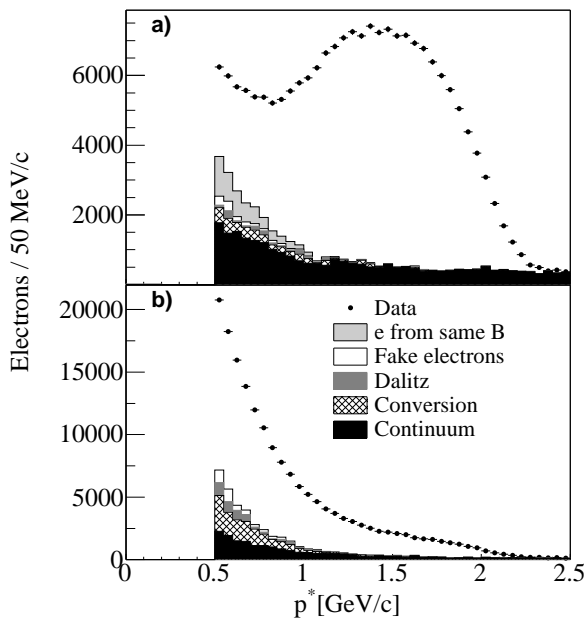} 
\caption{Measured momentum spectrum (points) and estimated backgrounds 
(histograms) for electron candidates in (a) the 
unlike-sign sample, and (b) the like-sign sample. } 
\label{rawspectra}
\end{center}
\end{figure}

Fig.~\ref{rawspectra} shows the electron momentum spectra and the background 
contributions discussed so far.  
Further backgrounds arise from decays of $\tau$ leptons, charmed 
mesons produced in $b \ra c\cbar s$ decays and $\jpsi$ or 
$\psitwos \to \epem$ decays with only one detected $e$. We also need to correct
for cases where the tagged electron does not originate from a semileptonic $B$ decay.
These backgrounds are irreducible, and their contributions to the three electron 
samples are estimated from MC simulations, using the ISGW2 model~\cite{isgw2} to describe 
semileptonic $D$ and $D_s$ meson decays. Assuming 
$\Gamma(D_s\to Xe\nu)=\Gamma(D\to Xe\nu)$, 
we obtain $\BR(D_s \to X e \nu) = (8.05 \pm 0.66)\%$.
Using $0.84\pm0.09$ \cite{cerncombined2001} for the measured fraction of 
$B \to D_s X $ decays where the $D_s$ originates from fragmentation 
of the $W$ Boson, and  $\BR (B \to D_s X ) = (10.5 \pm 2.6) \%$~\cite{PDG03} yields  
$\BR(B^{0,+} \to D_s^+ \to e^+) = (0.71 \pm 0.20)\%$. 
Assuming equal production rates of  $D$ and $ D^{*}$ and using
$\BR(\B \to \overline{D} D^{(*)} X)=(8.2 \pm 1.3)\%$~\cite{cerncombined2001},
we arrive at $\BR(B^{0,+} \to D^{0,+} \to e^+) = (0.84 \pm 0.21)\%$. 
To estimate the contribution of electrons from $\tau$ decays,
we consider the cascades $B \to \tau \to e$ and 
$B \to D_s \to \tau \to e$, with branching fractions taken from~\cite{PDG03}.
The rates for the decays $\B \to \jpsi \to e^+ e^-$ and $\B \to \psitwos \to e^+ e^-$
are also adjusted to~\cite{PDG03}.

These irreducible background spectra are subtracted from the like-sign
and unlike-sign spectra after correction for electron identification
efficiency.  We determine this efficiency as a function of $p^*$ and the
polar angle $\theta^*$ using $e^+e^-\to e^+e^-\gamma$ events and then use 
MC simulation to estimate losses in hadronic events with higher 
multiplicities. For $p^* > 0.6 \gevc$, the average efficiency is 91\% with 
an uncertainty of 1.5\% estimated from the size of the MC correction.
A summary of the yields is given in Table~\ref{grandtable}.

\begin{table}[h]
\begin{center}
\caption{Unlike-sign and like-sign pair yields for $0.6 < p^* < 2.5 \gevc$  
and their corrections with statistical and systematic errors.  
Numbers are quoted after all selection criteria.}
\begin{tabular}{lr@{\,}c@{\,}r@{\,}c@{\,}rr@{\,}c@{\,}r@{\,}c@{\,}r}
\hline \hline
 &\multicolumn{5}{c}{$e^+e^-$ sample } &\multicolumn{5}{c}{$e^{\pm} e^{\pm}$ sample }  \\
\hline 
All candidates &183493 &$\pm$ &434 & & &133842 &$\pm$ &371 & &  \\
continuum bkgd. &22922 &$\pm$ &349 & & &15758 &$\pm$ &290 & &  \\
conversion, Dalitz &2978 &$\pm$ &286 &$\pm$ &327 &10730 &$\pm$ &502 &$\pm$ &1177  \\
fake $e$ &885 &$\pm$ &63 &$\pm$ &423 &2229 &$\pm$ &182 &$\pm$ &966  \\
$e$ from same $B$ &3200 &$\pm$ &34 &$\pm$ &160 & & & & &  \\
\hline 
$e$ yield &153508 &$\pm$ &630 &$\pm$ &558 &105126 &$\pm$ &712 &$\pm$ &1523  \\
eff. corr. $e$ yield &169654 &$\pm$ &732 &$\pm$ &2235 &117192 &$\pm$ &803 &$\pm$ &2510  \\
\hline 
irreducible bkgd. &13912 &$\pm$ &92 &$\pm$ &1341 &14512 &$\pm$ &97 &$\pm$ &2513  \\
\hline 
corr. $e$ yield &155742 &$\pm$ &738 &$\pm$ &2606 &102680 &$\pm$ &809 &$\pm$ &3551  \\
\hline \hline
\end{tabular} 
\label{grandtable} 
\end{center}
\end{table}

To account for \BzBzb\ mixing, we determine the number of primary electrons in 
the $i$-th \pstar\ bin from the like-sign and unlike-sign pairs as
\begin{equation}
N^i_{b \to c,u}= \frac{1-f_0 \chi_0 }{1-2f_0 \chi_0 } \frac{N_{\uls}^i}{\epsilon_{\alpha^*}^i}
-  \frac{f_0 \chi_0 }{1-2f_0 \chi_0 } N_{\ls}^i
\label{eq:prompt}
\end{equation}
where $\chi_0 = 0.186 \pm 0.004$~\cite{PDG03} 
is the \BzBzb\ mixing parameter  and 
$f_0=\BR(\FourS \to \BzBzb)$ = $0.490\pm 0.018$~\cite{PDG03}.
The parameter $\epsilon_{\alpha^*}^i$ is the efficiency of the additional
requirement for the  unlike-sign sample as defined in Eq.~\ref{oacut}.

\begin{figure}[h]
\begin{center}
\includegraphics[height=1.8in]{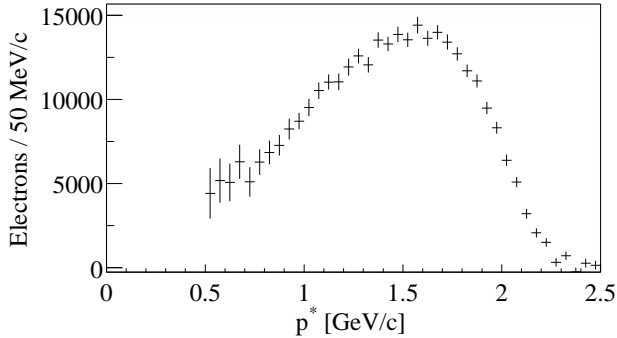} 
\caption {Electron momentum spectrum from $\B \to X e \nu (\gamma)$ decays in the
$\Upsilon(4S)$ frame after correction for efficiencies and bremsstrahlung, with 
combined statistical and systematic errors.} 
\label{fig_spectrum}
\end{center}
\end{figure}

The spectrum obtained from Eq.~\ref{eq:prompt} is corrected for the 
effects of bremsstrahlung in the detector material using MC simulation.
Since this correction significantly impacts the first moments, 
3\% for $\emin=0.6 \gev$ and 0.5\% for $\emin=1.5 \gev$, 
we have verified that the detector material is simulated to better 
than 3\%. Fig.~\ref{fig_spectrum} shows the resulting spectrum of primary 
electrons.

\begin{figure*}
\begin{center}
\includegraphics[height=1.8in]{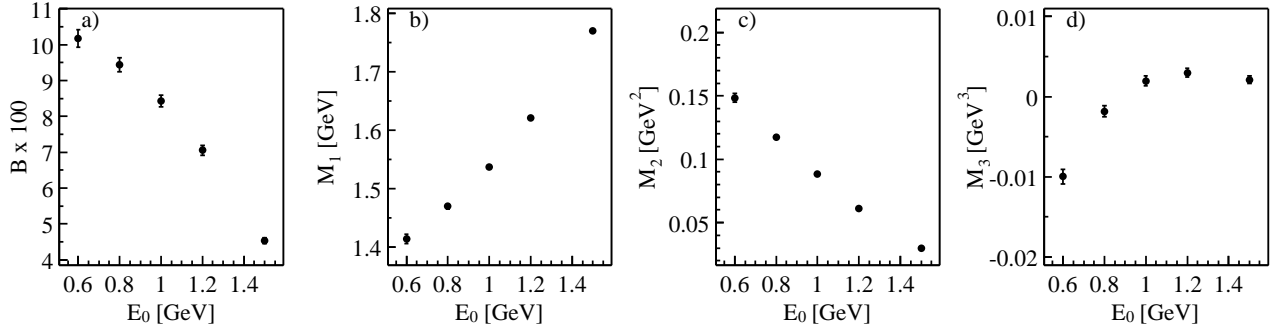} 
\caption{Measured moments of the inclusive electron energy spectrum of  $\B \to X_c e \nu (\gamma)$ decays as a function of the cut-off energy,
(a) ${\cal B}$, (b) $M_1$, (c) $M_2$ and (d) $M_3$. }
\label{fig:mom}
\end{center}
\end{figure*}

Charmless semileptonic \mbtoxu\ decays are modeled as 
in~\cite{btou} by a combination of semileptonic decays  
with resonant and non-resonant hadronic systems.
Using ${\cal B}(B \to X_u e \nu)  = (2.2 \pm 0.5) \times 10^{-3}$~\cite{btou} 
to correct for this background, we determine the moments  
${\tilde{M_n}}=\sum_k p_k^n N^k_{b\to c}/\sum_k  N^k_{b \to c}	$
where $k$ runs over all bins above the energy \emin 
and $p_k$ are the bin centers for $n=1$ and the bin centers shifted by 
${\tilde{M}}_1$ for $n=2,3$. These moments are then transformed into $E_e$
moments $M_n$ by correcting for the movement of the \B\ mesons in the 
center-of-mass frame. Further biases due to the event selection criteria and
binning are estimated from MC simulation.
The spectra and moments presented are those of  $\B \to X_c e \nu (\gamma)$
decays with any number of photons. The moments as a function of \emin are shown 
in Fig.~\ref{fig:mom} and Table~\ref{tab:sys} lists the principal systematic 
errors for \emin = 0.6 and 1.5 \gev. Without subtraction of \mbtoxu\ decays, we measure
$M_1^{b \to x}(1.5 \gev) = (1779.0 \pm 1.9 \pm 0.7) \mev$, which is consistent
with a recent measurement by CLEO~\cite{m1cleo}. 
Measurements with $E_0=0 \gev$ have been performed by DELPHI~\cite{m1delphi}.

We determine the partial branching fraction as
$(\sum_{k}N^k_{b\to c,u})/(N_{tag}\; \epsilon_{evt}\;\epsilon_{cuts})$, 
where $k$ runs over all bins with $E_e > \emin$, 
$N_{tag}=(3616.8$ $\pm$ 3.5(stat.) $\pm$ 21.8(syst.)) $\times 10^3$ 
is the background-corrected number of tag electrons, 
$\epsilon_{evt}=(98.9 \pm 0.5)\%$ refers to the relative efficiency for selecting 
two-electron events compared to events with a single $e_{tag}$, and
$\epsilon_{cuts}=(82.8\pm 0.3)\%$ is the acceptance for the signal electron for 
$\emin=0.6\gev$. The result,
\begin{equation*}
\begin{aligned}
\BR(\B\to X e \nu (\gamma), & \;E_e>0.6\gev) & \\
& = (10.36 \pm 0.06({\rm stat.}) \pm 0.23({\rm syst.}))\%, 
\end{aligned}
\end{equation*}
is consistent with our previous measurement~\cite{oldprl}, with the overall error
improved by 25\%. The partial branching fraction can be extrapolated to 
$\emin=0$ as part of a combined fit of the Heavy Quark Effective Theory (HQET) parameters 
to the full set of moments~\cite{comb}.

Current theoretical predictions on the lepton energy moments do not incorporate 
photon emission. Therefore we use PHOTOS~\cite{photos} to simulate QED radiation
and correct the moments for its impact. We verify that radiation that is not 
included in PHOTOS, e.g. additional hard photons, have no significant effect on 
the moments. The radiatively corrected moments and the estimated PHOTOS \
uncertainty~\cite{photos2} are given in Table~\ref{tab:sys}. The complete listing 
of all moments and the full correlation matrix, with and without PHOTOS corrections
can be found in Tables~\ref{tab:epaps-1}-\ref{tab:allcov}. For fitting purposes, a 
set of tables and matrices with a precision of 5 significant digits can be obtained 
from the authors.

In summary, we report a measurement of the electron energy spectrum of the inclusive 
decay \mbtox\  and its branching fraction for electron energies above 0.6 \gev, which
supersedes our previous result~\cite{oldprl}. We have also derived branching fractions, 
first, second, and third moments of electron energy spectrum from \mbtoxc\  decays for
cut-off energies from 0.6 to 1.5 \gev. This set of moments combined with hadron 
mass moments~\cite{hadmom} will be used for a significantly improved determination of 
HQET parameters and of $|V_{cb}|$~\cite{comb}.


\begin{table*}[ht] 
\caption{Results and breakdown of the systematic errors for 
$\BR = \tau_B \int_{\emin}^{\infty} (d\Gamma/dE_e)\,dE_e$  , and the moments $M_1$, $M_2$, and $M_3$
for $B\to X_c e\nu$ in the $B$-meson rest frame for two values of \emin. } 
\begin{center} 
\begin{tabular}{lcccccccc}
\hline 
\hline 
 &\multicolumn{2}{c}{${\cal B} [10^{-2}]$} &\multicolumn{2}{c}{$M_1 [\mev]$} &\multicolumn{2}{c}{$M_2 [10^{-3} \gev^2]$} &\multicolumn{2}{c}{$M_3 [10^{-3} \gev^3]$}  \\
$E_{0} [\gev]$ &0.6 &1.5 &0.6 &1.5 &0.6 &1.5 &0.6 &1.5  \\
\hline 
conversion and Dalitz pairs &$0.029$ &$0.001$ &$1.6$ &$0.02$ &$0.6$ &$0.00$ &$0.06$ &$0.00$  \\
$e$ identification efficiency &$0.151$ &$0.044$ &$2.5$ &$0.30$ &$0.6$ &$0.07$ &$0.29$ &$0.08$  \\
$e$ from same $B$ &$0.019$ &$0.000$ &$1.3$ &$0.00$ &$0.6$ &$0.00$ &$0.03$ &$0.00$  \\
$\B \to D_s \to e$ &$0.074$ &$0.001$ &$4.1$ &$0.04$ &$1.6$ &$0.00$ &$0.14$ &$0.00$  \\
$\B \to D \to e$ &$0.060$ &$0.000$ &$3.8$ &$0.00$ &$1.6$ &$0.00$ &$0.01$ &$0.00$  \\
$\B \to \tau \to e$ &$0.032$ &$0.002$ &$1.4$ &$0.05$ &$0.4$ &$0.00$ &$0.13$ &$0.00$  \\
$e$ from $J/\psi$ or $\psi(2S)$ &$0.002$ &$0.001$ &$0.0$ &$0.01$ &$0.0$ &$0.01$ &$0.00$ &$0.00$  \\
Secondary tags &$0.053$ &$0.011$ &$1.5$ &$0.06$ &$0.5$ &$0.00$ &$0.06$ &$0.00$  \\
$\chi$ &$0.034$ &$0.021$ &$0.8$ &$0.01$ &$0.3$ &$0.00$ &$0.03$ &$0.00$  \\
tracking efficiency &$0.084$ &$0.033$ &$1.0$ &$0.06$ &$0.3$ &$0.02$ &$0.07$ &$0.00$  \\
bremsstrahlung correction &$0.011$ &$0.028$ &$1.9$ &$0.43$ &$0.0$ &$0.05$ &$0.19$ &$0.00$  \\
event selection &$0.052$ &$0.024$ &$0.6$ &$0.14$ &$0.0$ &$0.03$ &$0.07$ &$0.01$  \\
$b \to u$ subtraction  &$0.047$ &$0.030$ &$1.2$ &$1.24$ &$0.6$ &$0.48$ &$0.20$ &$0.17$  \\
$B$ momentum correction &$0.000$ &$0.005$ &$0.0$ &$0.19$ &$0.1$ &$0.10$ &$0.04$ &$0.02$  \\
$N_{tag}$ normalization &0.068 &0.030 & & & & & &  \\
\hline 
Moments &$10.17$ &$4.54$ &$1414.3$ &$1769.2$ &$148.5$ &$29.8$ &$-9.97$ &$2.11$  \\
$\pm$(stat.) &$0.06$ &$0.03$ &$3.7$ &$1.8$ &$2.0$ &$0.8$ &$0.79$ &$0.44$  \\
$\pm$(sys.) &$0.23$ &$0.08$ &$7.4$ &$1.4$ &$2.7$ &$0.5$ &$0.48$ &$0.20$  \\
\hline 
Moments with rad. correction &$10.30$ &$4.79$ &$1432.8$ &$1774.3$ &$148.0$ &$30.3$ &$-12.05$ &$2.12$  \\
$\pm$(stat.) &\ $0.06$\  &\ $0.03$\  &\ $3.9$\  &\ $1.9$\  &\ $2.2$\  &\ $0.9$\  &\ $0.88$\  &\ $0.47$\   \\
$\pm$(sys.) &\ $0.24$\  &\ $0.09$\  &\ $7.8$\  &\ $1.4$\  &\ $3.1$\  &\ $0.5$\  &\ $0.46$\  &\ $0.20$\   \\
\hline 
\hline 
\end{tabular} 

\label{tab:sys}
\end{center} 
\end{table*}

We are grateful for the excellent luminosity and machine conditions
provided by our \pep2\ colleagues, 
and for the substantial dedicated effort from
the computing organizations that support \babar.
The collaborating institutions wish to thank 
SLAC for its support and kind hospitality. 
This work is supported by
DOE
and NSF (USA),
NSERC (Canada),
IHEP (China),
CEA and
CNRS-IN2P3
(France),
BMBF and DFG
(Germany),
INFN (Italy),
FOM (The Netherlands),
NFR (Norway),
MIST (Russia), and
PPARC (United Kingdom). 
Individuals have received support from the 
A.~P.~Sloan Foundation, 
Research Corporation,
and Alexander von Humboldt Foundation.

\begin{table*}[!hb]
\begin{center}
\caption{Measured moments $M_1$,  $M_2$,  $M_3$, and ${\cal B}$
for five cut-off energies $E_0$ with first their statistical error and 
second their systematic error. The moments are given in the $B$-meson rest 
frame and are defined to include $X_c$ hadronic states only 
and to include decays $B \to X_c e \nu \gamma$ with any number of photons.
}

\begin{tabular}{crcrcrrcrcrrcrcrrcrcr}
\hline \hline
$E_{0} [\gev] $ &\multicolumn{5}{c}{${\cal B} [10^{-2}]$ } &\multicolumn{5}{c}{$M_1 [\mev]$ } &\multicolumn{5}{c}{$M_2 [10^{-3} \gev^2]$ } &\multicolumn{5}{c}{$M_3 [10^{-3} \gev^3]$ }  \\
\hline 
0.6 &\ \ $10.17$ &$\pm$ &$0.06$ &$\pm$ &$0.23$ &\ \ $1414.3$ &$\pm$ &$3.7$ &$\pm$ &$7.4$ &\ \ $148.5$ &$\pm$ &$2.0$ &$\pm$ &$2.7$ &\ \ $-9.97$ &$\pm$ &$0.79$ &$\pm$ &$0.48$  \\
0.8 &\ \ $9.43$ &$\pm$ &$0.05$ &$\pm$ &$0.19$ &\ \ $1469.8$ &$\pm$ &$2.4$ &$\pm$ &$3.6$ &\ \ $117.3$ &$\pm$ &$1.2$ &$\pm$ &$0.9$ &\ \ $-1.82$ &$\pm$ &$0.60$ &$\pm$ &$0.39$  \\
1.0 &\ \ $8.42$ &$\pm$ &$0.04$ &$\pm$ &$0.16$ &\ \ $1537.2$ &$\pm$ &$1.9$ &$\pm$ &$2.1$ &\ \ $88.3$ &$\pm$ &$0.9$ &$\pm$ &$0.6$ &\ \ $1.95$ &$\pm$ &$0.53$ &$\pm$ &$0.30$  \\
1.2 &\ \ $7.05$ &$\pm$ &$0.04$ &$\pm$ &$0.13$ &\ \ $1621.3$ &$\pm$ &$1.8$ &$\pm$ &$1.6$ &\ \ $61.2$ &$\pm$ &$0.9$ &$\pm$ &$0.6$ &\ \ $2.97$ &$\pm$ &$0.49$ &$\pm$ &$0.24$  \\
1.5 &\ \ $4.54$ &$\pm$ &$0.03$ &$\pm$ &$0.08$ &\ \ $1769.2$ &$\pm$ &$1.8$ &$\pm$ &$1.4$ &\ \ $29.8$ &$\pm$ &$0.8$ &$\pm$ &$0.5$ &\ \ $2.11$ &$\pm$ &$0.44$ &$\pm$ &$0.20$  \\
\hline \hline
\end{tabular} 

\label{tab:epaps-1}
\end{center} 
\end{table*}

\begin{table*}[!hb]
\begin{center}
\caption{Measured moments $M_1$, $M_2$, $M_3$, and ${\cal B}$
for five cut-off energies $E_0$ with first their statistical error and 
second their systematic error.The moments are given in the $B$-meson rest frame and 
are corrected for QED radiative effects.}


\begin{tabular}{crcrcrrcrcrrcrcrrcrcr}
\hline \hline
$E_{0} [\gev] $ &\multicolumn{5}{c}{${\cal B} [10^{-2}]$ } &\multicolumn{5}{c}{$M_1 [\mev]$ } &\multicolumn{5}{c}{$M_2 [10^{-3} \gev^2]$ } &\multicolumn{5}{c}{$M_3 [10^{-3} \gev^3]$ }  \\
\hline 
0.6 &\ \ $10.30$ &$\pm$ &$0.06$ &$\pm$ &$0.24$ &\ \ $1432.8$ &$\pm$ &$3.9$ &$\pm$ &$7.8$ &\ \ $148.0$ &$\pm$ &$2.2$ &$\pm$ &$3.1$ &\ \ $-12.05$ &$\pm$ &$0.88$ &$\pm$ &$0.46$  \\
0.8 &\ \ $9.61$ &$\pm$ &$0.05$ &$\pm$ &$0.20$ &\ \ $1484.8$ &$\pm$ &$2.6$ &$\pm$ &$3.7$ &\ \ $117.7$ &$\pm$ &$1.3$ &$\pm$ &$1.0$ &\ \ $-3.18$ &$\pm$ &$0.64$ &$\pm$ &$0.38$  \\
1.0 &\ \ $8.65$ &$\pm$ &$0.04$ &$\pm$ &$0.17$ &\ \ $1548.7$ &$\pm$ &$2.0$ &$\pm$ &$2.2$ &\ \ $89.1$ &$\pm$ &$1.0$ &$\pm$ &$0.6$ &\ \ $1.19$ &$\pm$ &$0.57$ &$\pm$ &$0.30$  \\
1.2 &\ \ $7.31$ &$\pm$ &$0.04$ &$\pm$ &$0.14$ &\ \ $1629.9$ &$\pm$ &$1.9$ &$\pm$ &$1.7$ &\ \ $62.1$ &$\pm$ &$0.9$ &$\pm$ &$0.6$ &\ \ $2.66$ &$\pm$ &$0.52$ &$\pm$ &$0.24$  \\
1.5 &\ \ $4.79$ &$\pm$ &$0.03$ &$\pm$ &$0.09$ &\ \ $1774.3$ &$\pm$ &$1.9$ &$\pm$ &$1.4$ &\ \ $30.3$ &$\pm$ &$0.9$ &$\pm$ &$0.5$ &\ \ $2.12$ &$\pm$ &$0.47$ &$\pm$ &$0.20$  \\
\hline \hline
\end{tabular} 

\label{tab:epaps-2}
\end{center} 
\end{table*}

\clearpage \newpage

\begin{sidewaystable}[h]
\caption{
Correlation matrix of the 20 measured moments. Matrix elements are
for the full errors (statistical and systematic added in quadrature)
in percent. 
The superscript $i$ refers to the energy thresholds
$E_i=0.6$, 0.8, 1.0, 1.2, 1.5 GeV for $i=1$, 2, 3, 4, 5 respectively. \\[5mm]}
\small

\begin{tabular}{crrrrrrrrrrrrrrrrrrrr}
\hline 
\hline 
  &$M_1^{1}$ &$M_1^{2}$ &$M_1^{3}$ &$M_1^{4}$ &$M_1^{5}$ &$M_2^{1}$ &$M_2^{2}$ &$M_2^{3}$ &$M_2^{4}$ &$M_2^{5}$ &$M_3^{1}$ &$M_3^{2}$ &$M_3^{3}$ &$M_3^{4}$ &$M_3^{5}$ &${\cal B}^{1}$ &${\cal B}^{2}$ &${\cal B}^{3}$ &${\cal B}^{4}$ &${\cal B}^{5}$  \\
$M_1^{1}$ &100.0 &83.2 &64.0 &48.3 &34.0 &-73.6 &-26.5 &12.3 &22.5 &23.4 &13.5 &-9.2 &0.7 &10.5 &15.7 &-50.8 &-26.4 &-11.8 &-2.5 &8.0  \\
$M_1^{2}$ & &100.0 &81.4 &66.8 &51.5 &-35.1 &-20.0 &27.3 &37.8 &38.3 &-7.3 &-0.3 &9.7 &21.3 &27.2 &-36.5 &-24.1 &-7.4 &3.3 &16.7  \\
$M_1^{3}$ & & &100.0 &84.0 &68.5 &-3.5 &26.6 &40.4 &53.9 &54.5 &5.5 &6.6 &21.0 &34.3 &41.1 &-16.7 &-7.2 &-0.8 &12.0 &27.8  \\
$M_1^{4}$ & & & &100.0 &79.1 &15.5 &49.0 &69.8 &61.9 &65.7 &23.8 &28.0 &33.9 &43.2 &51.6 &-0.3 &6.7 &12.1 &15.9 &35.1  \\
$M_1^{5}$ & & & & &100.0 &28.7 &63.1 &85.6 &88.8 &79.7 &48.7 &59.7 &64.4 &66.4 &65.1 &17.6 &23.4 &27.5 &31.5 &36.3  \\
$M_2^{1}$ & & & & & &100.0 &66.5 &42.8 &35.2 &33.4 &3.9 &41.9 &38.0 &33.4 &30.9 &66.1 &44.1 &33.9 &29.6 &28.0  \\
$M_2^{2}$ & & & & & & &100.0 &75.2 &68.9 &67.3 &59.0 &61.0 &63.1 &61.4 &60.6 &53.9 &51.4 &40.3 &38.0 &41.1  \\
$M_2^{3}$ & & & & & & & &100.0 &90.9 &88.6 &66.7 &81.3 &79.7 &78.9 &78.8 &36.3 &39.4 &41.9 &39.3 &45.1  \\
$M_2^{4}$ & & & & & & & & &100.0 &94.5 &66.3 &82.1 &87.4 &87.3 &84.9 &30.3 &35.3 &38.4 &41.9 &44.0  \\
$M_2^{5}$ & & & & & & & & & &100.0 &67.4 &83.8 &90.8 &94.5 &93.6 &29.0 &34.3 &37.3 &40.1 &45.3  \\
$M_3^{1}$ & & & & & & & & & & &100.0 &79.0 &76.8 &72.2 &68.2 &31.4 &43.7 &41.6 &38.2 &34.4  \\
$M_3^{2}$ & & & & & & & & & & & &100.0 &95.9 &90.5 &85.2 &47.5 &48.3 &48.5 &44.5 &39.5  \\
$M_3^{3}$ & & & & & & & & & & & & &100.0 &97.4 &92.8 &41.4 &44.0 &44.8 &44.0 &40.0  \\
$M_3^{4}$ & & & & & & & & & & & & & &100.0 &97.5 &33.7 &37.4 &39.1 &40.6 &39.5  \\
$M_3^{5}$ & & & & & & & & & & & & & & &100.0 &28.3 &32.4 &34.5 &36.1 &39.3  \\
${\cal B}^{1}$ & & & & & & & & & & & & & & & &100.0 &95.4 &90.4 &86.3 &80.1  \\
${\cal B}^{2}$ & & & & & & & & & & & & & & & & &100.0 &97.7 &95.1 &90.1  \\
${\cal B}^{3}$ & & & & & & & & & & & & & & & & & &100.0 &98.3 &94.3  \\
${\cal B}^{4}$ & & & & & & & & & & & & & & & & & & &100.0 &96.5  \\
${\cal B}^{5}$ & & & & & & & & & & & & & & & & & & & &100.0  \\
\hline 
\hline 
\end{tabular} 

\label{tab:allcov}
\end{sidewaystable}

\clearpage \newpage


\end{document}